\documentstyle[preprint,prc,aps]{revtex}
\begin{document}
\title{Important configurations for $NN$ processes in a Goldstone boson
exchange model}
\author{D. Bartz\thanks{e-mail : d.bartz@ulg.ac.be}
and Fl. Stancu\thanks{e-mail : fstancu@ulg.ac.be}}
\address{Universit\'{e} de Li\`ege, Institut de Physique B.5, Sart Tilman,
B-4000 Li\`ege 1, Belgium}
\date{\today}
\maketitle
\everymath={\displaystyle}

\vspace{1cm}

\begin{abstract}
We study the short-range nucleon-nucleon interaction in a nonrelativistic
chiral constituent quark model by diagonalizing a Hamiltonian
containing a linear confinement and a Goldstone boson exchange
interaction between quarks. A finite six-quark basis obtained
from single particle cluster model states
was previously used. Here
we show that the configurations which appear naturally through the use of
molecular orbitals, instead of cluster model states,
are much more efficient in lowering the six-quark
energy.\par
\end{abstract}

\section{Introduction}
Constituent quark models have been applied to the study of the nucleon-nucleon
interaction. In a category of such models the 
Hamiltonian contains a kinetic term, a
confinement term and an effective one-gluon exchange (OGE) term. These models
explain the short-range repulsion in the $NN$ systems as due to the
colour-magnetic part of the OGE interaction combined with quark interchanges
between the $3q$ clusters. Nevertheless, an effective meson-exchange potential,
introduced through the coupling of mesons to $3q$ cluster collectively, was
required in order to reproduce the intermediate- and long-range attraction
(for a review see for example
\cite{OK84,MY88,SH89}).\par
Another category are the hybrid models \cite{KU91,ZH94,FU96}.
There, in addition to the
OGE interaction, the quarks belonging to different $3q$
clusters interact via pseudoscalar and scalar meson exchange. In these models
the short-range repulsion in the $NN$ system is still attributed to the OGE
interaction between the constituent quarks. The medium- and long-
range attraction are due to meson exchange, as expected. \par 
In a recent exploratory work \cite{ST97}, by using the Born-Oppenheimer
approximation, we calculated an effective $NN$ interaction at zero separation
distance, within the constituent quark model \cite{GL96a,GL96b,GL97}. In
this model the quarks interact via Goldstone boson exchange (GBE)
instead of OGE of conventional models, and the
hyperfine splitting in hadrons is obtained from the short-range part of the GBE
interaction. An important merit of the GBE model is that it reproduces the
correct order of positive and negative parity states in both nonstrange
\cite{GL96b} and strange baryons \cite{GL97}
in contrast to any OGE model. 
In Ref. \cite{ST97} we showed
that the same short-range part of the GBE interaction, also induces a
short-range repulsion in the $NN$ system. Moreover, the long and
middle range attraction of the $NN$ potential will automatically appear due to 
the presence of a Yukawa potential tail in the qq interaction
and due to $2\pi$ (or sigma) exchanges. \par
In \cite{ST97} the
height of the repulsive core was about 800 MeV for the $^3S_1$ channel and
1300 MeV for the $^1S_0$ channel. Such a result has been obtained from
diagonalizing the Hamiltonian of Ref. \cite{GL96b} in a six-quark
cluster model basis built from harmonic oscillator states
containing up to two quanta of excitation.
The six-quark states have orbital symmetries $[6]_O$ and $[42]_O$,
so that they contain configurations
of type $s^6$, $s^4p^2$ 
and $s^52s$, with the centre of mass motion removed.
In the flavour-spin space only
the symmetries $[33]$, $[51]$ and $[411]$ were retained.
As shown in \cite{ST97} they produce the most important
five basis states allowed by the Pauli principle. Due
to the specific flavour-spin structure of the GBE interaction, we found
that the
state $|s^4p^2[42]_O[51]_{FS}\rangle$ was highly dominant at zero-separation
between nucleons. The symmetry structure of this state implies the existence of
a node in the nucleon-nucleon
$S$-wave relative motion wave function at short distances.
This nodal structure will induce an additional effective repulsion
in dynamical calculations based, for example, on the resonating
group method.\par
A central issue of the $NN$ problem is the construction of an adequate
six-quark
basis states. In principle the choice of basis is arbitrary if a sufficiently
large basis is considered in the Hamiltonian diagonalization. But, as in
practice one considers a finite set, its choice is very important. Ref.
\cite{ST87} advocated the use of molecular-type single particle orbitals
instead of cluster model-type states. These orbitals have the proper axially
and reflectionally symmetries and can be constructed from appropriate
combinations of two-centre Gaussians. At zero-separation the six-quark states
obtained from such orbitals contain certain $p^ns^{6-n}$ components which are
missing in the cluster model basis. In Ref. \cite{ST88} it has been shown
that for an OGE model used in the calculations of the $NN$ potential
they lead to a substantial lowering of the lowest eigenstate,
used in the calculation of the $NN$ potential. The molecular orbitals have also
the advantage of forming an orthogonal and complete basis while the cluster
model (two-centre) states are not orthogonal and are overcomplete.\par
Due to the predominance ( 93 \% ) of only one component, namely 
$|s^4p^2[42]_O[51]_{FS}\rangle$, in the ground state wave function obtained
in a
cluster model basis \cite{ST97} the GBE model is a more chalenging case to
test the efficiency of a molecular orbital basis than the OGE model, where
there is some mixture of states (see e.g. \cite{OK84,ST88}). Here we show 
that by using molecular orbitals the height of the repulsion reduces by about 
22 \% and 25 \% in the $^3S_1$ and $^1S_0$
channels respectively. \par 
The paper is organized as follows. In Sec. 2 we briefly recall the procedure of
constructing six-quark states from molecular orbital single particle states.
In Sec. 3 we describe the GBE Hamiltonian \cite{GL96b}. In Sec. 4 we present
our results for zero-separation $NN$ interaction derived in the
Born-Oppenheimer
approximation for the $IS$ = (10) and (01) sectors. The last section
is devoted to summary and conclusions.
\section{Six-quark states from molecular orbitals}
Here we follow closely Ref. \cite{ST87} where the use of molecular
orbitals in the construction of six-quark states was originally proposed,
instead of commonly used cluster model states. Let us denote by $Z$
the separation coordinate between the centres of the two clusters. 
At finite $Z$, in the
simplest cluster model basis, each of the six quarks is decribed by
an orbital wave function represented by a Gaussian centered either
at $Z/2$ or $-Z/2$. These nonorthogonal states are denoted by $R$ (right) and
$L$ (left) respectively
\begin{eqnarray}
&R(\vec{r}) = \Psi(\vec{r} - \vec{Z}/2),\,\,\,\,\,\,
 L(\vec{r}) = \Psi(\vec{r} + \vec{Z}/2) .&
\end{eqnarray}
\par
Alternatively, in a molecular basis we consider the two lowest states,
$\sigma$ which is even and $\pi$ which is
odd. These could be either the solutions
of a static, axially and reflectionally symmetric independent particle
model Hamiltonian (see for example \cite{KO94}) or, as for the present
purpose, can be constructed from $R$ and $L$ states.\par
First we introduce pseudo-right and pseudo-left states $r$ and $l$
starting from the molecular orbitals $\sigma$ and $\pi$ as
\begin{equation}
\left[ \begin{array}{c} r\\ l \end{array} \right] =
2^{-1/2}\, ( \sigma \pm \pi ) \hspace{5mm} 
\rm{for\ all\ }Z,
\end{equation}
where
\begin{eqnarray}
& <r|r> = <l|l> = 1 ,\,\, <r|l> = 0 .&
\end{eqnarray}
On the other hand, starting from the cluster model states,
one can construct good parity, orthonormal states for all $Z$ by setting
\begin{equation}
\left[ \begin{array}{c} \sigma\\ \pi\end{array} \right] =
[2( 1 \pm <R|L>)]^{-1/2} ( R \pm L ) ,
\end{equation}
which, introduced in (2) gives
\begin{equation}
\left[ \begin{array}{c} r\\ l \end{array} \right] =
\frac{1}{2} \left[ \frac{R+L}{(1+<R|L>)^{1/2}} \pm
\frac{R-L}{(1-<R|L>)^{1/2}} \right].
\end{equation}
\par
At $Z \rightarrow 0$ one has
$\sigma \rightarrow s$ and $\pi \rightarrow p$ (with $m = 0,\pm1$)
, so that
\begin{equation}
\left[ \begin{array}{c} r\\ l \end{array} \right] =
2^{1/2} (s \pm p) ,
\end{equation}
and at $Z \rightarrow \infty$
 one has $r \rightarrow R$ and $l \rightarrow L$.\par
From $(r,l)$ as well as from $(\sigma,\pi)$ orbitals one can construct
six-quark states of required permutation symmetry. For the $S_6$
symmetries relevant for the $NN$ problem the transformations
between six-quark states expressed in terms of $(r,l)$ and $(\sigma,\pi)$
states are given in Table I of Ref. \cite{ST87}. This table shows that in
the limit $Z \rightarrow 0$ six-quark states obtained from molecular
orbitals contain configurations of type $s^np^{6-n}$ with $n = 0,1,...,
6$. For example the $[6]_O$ state contains $s^6$,
$s^6p^4$, $s^2p^4$ and $p^6$ configurations and the $[42]_O$ state
associated to
the $S$-channel contains $s^4p^2$ and $s^2p^4$ configurations. This is in
contrast to the cluster model basis where $[6]_O$ contains only $s^6$ and
$[42]_O$ only $s^4p^2$ configurations \cite{HA81}. This suggests that the
six-quark basis states constructed from molecular orbitals form a richer
basis without introducing more single particle states.
Here we examine its role in lowering the ground state energy of a
six-quark system described by the Hamiltonian introduced in the next
section.\par
Using Table I of Ref. \cite{ST87} we find that the six-quark basis states
needed for the $^3S_1$ or $^1S_0$ channels are:
\begin{eqnarray}
\left.{\left|{33{\left[{6}\right]}_{O}{\left[{33}\right]}_{FS}}\right.}
\right\rangle\
& = & \frac{1}{4}\ \left.{\left| \left[{\sqrt {5}\ \left({{s}^{6}\ -\
{p}^{6}}\right)\ -\
\sqrt {3}\ \left({{s}^{4}{p}^{2}\ -\ {s}^{2}{p}^{4}}\right)}\right]\
{{\left[{6}\right]}_{O}{\left[{33}\right]}_{FS}}\right.}\right\rangle, \\
\left.{\left|{33{\left[{42}\right]}_{O}{\left[{33}\right]}_{FS}}\right.}
\right\rangle\
& = & \sqrt {\frac{1}{2}}\ \left.{\left|{ \left[{{s}^{4}{p}^{2}\ -\
{s}^{2}{p}^{4}}\right]
{\left[{42}\right]}_{O}{\left[{33}\right]}_{FS}}\right.}\right\rangle, \\
\left.{\left|{33{\left[{42}\right]}_{O}{\left[{51}\right]}_{FS}}\right.}
\right\rangle\
& = & \sqrt {\frac{1}{2}}\ \left.{\left|{ \left[{{s}^{4}{p}^{2}\ -\
{s}^{2}{p}^{4}}\right]
{\left[{42}\right]}_{O}{\left[{51}\right]}_{FS}}\right.}\right\rangle, \\
\left.{\left|{33{\left[{42}\right]}_{O}{\left[{411}\right]}_{FS}}\right.}
\right\rangle\
& = & \sqrt {\frac{1}{2}}\ \left.{\left|{ \left[{{s}^{4}{p}^{2}\ -\
{s}^{2}{p}^{4}}\right]
{\left[{42}\right]}_{O}{\left[{411}\right]}_{FS}}\right.}\right\rangle, \\
\left.{\left|{{42}^{+}{\left[{6}\right]}_{O}{\left[{33}\right]}_{FS}}\right.}
\right\rangle\
& = & {\frac{1}{4}} \sqrt {{\frac{1}{2}}}\ \left.{\left|{ \left[{{\sqrt
{15}}^{}\left({{s}^{6}\ +\ {p}^{6}}\right)\ -\ \left({{s}^{4}{p}^{2}\ +\
{s}^{2}{p}^{4}}\right)}\right]{\left[{6}\right]}_{O}{\left[{33}\right]}_{FS}}
\right.}\right\rangle, \\
\left.{\left|{{42}^{+}{\left[{42}\right]}_{O}{\left[{33}\right]}_{FS}}\right.}
\right\rangle\
& = & \sqrt {\frac{1}{2}}\ \left.{\left|{ \left[{\left.{{s}^{4}
{p}^{2}}\right.\ +\
{s}^{2}{p}^{4}}\right]{\left[{42}\right]}_{O}{\left[{33}\right]}_{FS}}\right.}
\right\rangle, \\
\left.{\left|{{42}^{+}{\left[{42}\right]}_{O}{\left[{51}\right]}_{FS}}\right.}
\right\rangle\
& = & \sqrt {\frac{1}{2}}\ \left.{\left|{ \left[{\left.{{s}^{4}
{p}^{2}}\right.\ +\
{s}^{2}{p}^{4}}\right]{\left[{42}\right]}_{O}{\left[{51}\right]}_{FS}}\right.}
\right\rangle, \\
\left.{\left|{{42}^{+}{\left[{42}\right]}_{O}{\left[{411}\right]}_{FS}}\right.}
\right\rangle\
& = & \sqrt {\frac{1}{2}}\ \left.{\left|{ \left[{\left.{{s}^{4}
{p}^{2}}\right.\ +\
{s}^{2}{p}^{4}}\right]{\left[{42}\right]}_{O}{\left[{411}\right]}_{FS}}\right.}
\right\rangle, \\
\left.{\left|{{51}^{+}{\left[{6}\right]}_{O}{\left[{33}\right]}_{FS}}\right.}
\right\rangle\
& = & \frac{1}{4}\ \left.{\left|{ \left[{\sqrt {3}\ \left.{\left({{s}^{6}\ - \
{p}^{6}}\right)}\right.\ +\ \sqrt {5}\ \left({{s}^{4}{p}^{2}\ -\
{s}^{2}{p}^{4}}\right)}\right]{\left[{6}\right]}_{O}{\left[{33}\right]}_{FS}}
\right.}\right\rangle,
\end{eqnarray}
where the notation $33$ and $mn^+$ in the $lhs$ of each equality above means
$r^3\ell^3$ and $r^m\ell^n+r^n\ell^m$ as in Ref. \cite{ST87} (see also
discussion below). Each wave function contains an orbital part ($O$) and a
flavour-spin part ($FS$) which combined with the colour singlet $[222]_C$ state
gives rise to a totally antisymmetric state. We restricted the flavour-spin
states to $[33]_{FS}$, $[51]_{FS}$ and $[411]_{FS}$ according to the discussion
given in Sec. II of Ref. \cite{ST97} where the most important states have been
selected by using a schematic version of the Hamiltonian introduced
in the next section.\par
In a cluster model, the most important basis states built from $s$ and $p$
harmonic oscillator states are
\begin{equation}
\left.{\left|{{s}^{6}{\left[{6}\right]}_{O}{\left[{33}\right]}_{FS}}\right.}
\right\rangle,
\end{equation}
\begin{equation}
\left.{\left|{{s}^{4}{p}^{2}{\left[{42}\right]}_{O}{\left[{33}\right]}_{FS}}
\right.}\right\rangle,
\end{equation}
\begin{equation}
\left.{\left|{{s}^{4}{p}^{2}{\left[{42}\right]}_{O}{\left[{51}\right]}_{FS}}
\right.}\right\rangle,
\end{equation}
\begin{equation}
\left.{\left|{{s}^{4}{p}^{2}{\left[{42}\right]}_{O}{\left[{411}\right]}_{FS}}
\right.}\right\rangle.
\end{equation}
These are the first four states given by Eq. (8) of Ref. \cite{ST97}. The fifth
one, containing the configuraiton $s^52s$ is not considered here. Its role
in lowering the ground state energy by a few MeV 
proved to be negligible. 
Besides being poorer in $s^np^{6-n}$ configurations, as explained above, the
number of basis states is smaller in the cluster model although we deal
with the
same $[f]_O$ and $[f]_{FS}$ symmetries and the same harmonic oscillator states
$s$ and $p$ in both cases. This is due to the existence of three-quark clusters
only
in the cluster model states, while the molecular basis also allows
configurations with five quarks to the left and one to the right, or vice
versa, or four quarks to the left and two to the right or vice versa. At large
separations these states act as ``hidden colour" states but at zero separation
they bring a significant contribution, as we shall see below. \par 
The matrix elements of the Hamiltonian (22) are calculated in the basis (7-15)
by using the fractional parentage technique described in Refs. \cite{HA81,ST96}
and also applied in Ref. \cite{ST97}. A programme based on Mathematica 
\cite{MATH} has been
created for this purpose. In this way every six-body matrix element reduces
to a
linear combination of two-body matrix elements of either symmetric or
antisymmetric states for which Eqs. (3.3) of Ref. \cite{GL96a} can be used to
integrate in the spin-flavour space. Then the linear combinations contain 
orbital two-body matrix elements of the type $\left\langle{ss\left|{{V}_{\gamma
}}\right|ss}\right\rangle$, $\left\langle{ss\left|{{V}_{\gamma
}}\right|pp}\right\rangle$, $\left\langle{sp\left|{{V}_{\gamma
}}\right|sp}\right\rangle$, $\left\langle{sp\left|{{V}_{\gamma
}}\right|ps}\right\rangle$ and 
$\left\langle{pp\left|{{V}_{\gamma
}}\right|pp}\right\rangle_{L\ =\ 0}$ where $\gamma = \pi$, $\eta$ or $\eta '$,
see Eq. (25). Here we study the case $Z = 0$ for which the following
harmonic oscillator states are used
\begin{eqnarray}
|s> & = & \pi^{-3/4} \beta^{-3/2} \exp{(-r^2/2\beta^2)}, \\
|p> & = & 8^{1/2} 3^{-1/2} \pi^{-1/4} \beta^{-5/2} r \exp{(-r^2/2\beta^2)}
\, Y_{lm}\, .
\end{eqnarray}
In this basis the orbital two-body matrix elements of the linear confinement
$V_{conf} = Cr$ potential (23) are calculated analytically (see Appendix D
of Ref. \cite{ST97}).

\section{Hamiltonian}
The GBE Hamiltonian considered below has the form \cite{GL96b} :
\begin{equation}
H= \sum_i m_i + \sum_i \frac{\vec{p}_{i}^{\,2}}{2m_i} - \frac {(\sum_i
\vec{p}_{i})^2}{2\sum_i m_i} + \sum_{i<j} V_{\text{conf}}(r_{ij}) + \sum_{i<j}
V_\chi(r_{ij}) \, ,
\label{ham}
\end{equation}
with the linear confining interaction :
\begin{equation}
 V_{\text{conf}}(r_{ij}) = -\frac{3}{8}\lambda_{i}^{c}\cdot\lambda_{j}^{c} \, C
\, r_{ij} \, ,
\label{conf}
\end{equation}
and the spin--spin component of the GBE interaction in its $SU_F(3)$ form :
\begin{eqnarray}
V_\chi(r_{ij})
&=&
\left\{\sum_{F=1}^3 V_{\pi}(r_{ij}) \lambda_i^F \lambda_j^F \right.
\nonumber \\
&+& \left. \sum_{F=4}^7 V_{K}(r_{ij}) \lambda_i^F \lambda_j^F
+V_{\eta}(r_{ij}) \lambda_i^8 \lambda_j^8
+V_{\eta^{\prime}}(r_{ij}) \lambda_i^0 \lambda_j^0\right\}
\vec\sigma_i\cdot\vec\sigma_j,
\label{VCHI}
\end{eqnarray}
\noindent
with $\lambda^0 = \sqrt{2/3}~{\bf 1}$, where $\bf 1$ is the $3\times3$ unit
matrix. The interaction (24) contains $\gamma = \pi, K, \eta$ and $\eta '$
meson-exchange terms and the form of $V_{\gamma} \left(r_{ij}\right)$ is given
as the sum of two distinct contributions : a Yukawa-type potential containing
the mass of the exchanged meson and a short-range contribution of opposite
sign, the role of which is crucial in baryon spectroscopy. For a given meson
$\gamma$, the exchange potential is
\begin{equation}V_\gamma (r)=
\frac{g_\gamma^2}{4\pi}\frac{1}{12m_i m_j}
\{\theta(r-r_0)\mu_\gamma^2\frac{e^{-\mu_\gamma r}}{ r}- \frac {4}{\sqrt {\pi}}
\alpha^3 \exp(-\alpha^2(r-r_0)^2)\}.
\end{equation}
For a system of $u$ and $d$ quarks only, as it is the case here,
the $K$-exchange does not contribute. In the calculations below we use the
parameters of Refs.\cite{GL96b}. These are :
\begin{eqnarray}
&\frac{g_{\pi q}^2}{4\pi} = \frac{g_{\eta q}^2}{4\pi}
= 0.67,\,\,
\frac{g_{\eta ' q}^2}{4\pi} = 1.206 ,&\nonumber\\
&r_0 = 0.43 \, { fm}, ~\alpha = 2.91 \, { fm}^{-1},~~
 C= 0.474 \, { fm}^{-2}, \, m_{u,d} = 340 \, { MeV}, \, &\\
&\mu_{\pi} = 139 \, { MeV},~ \mu_{\eta} = 547 \, { MeV},~
\mu_{\eta'} = 958 \, { MeV}.&
\nonumber
\end{eqnarray}
In principle it would be better to use a parametrization of the GBE 
interaction as given in \cite{GPVW98} based on a semirelativistic 
Hamiltonian. However, in applying the quark cluster approach to 
two-baryon systems we are restricted to use a nonrelativistic
kinematics and an $s^3$ wave function for the ground state baryon.
With an $s^3$ variational solution the nonrelativistic Hamiltonian
introduced above works generally well \cite{PS98}.
In particular, for the nucleon,
the quantity $< N | H | N >$ reaches its minimum at 969.6 MeV 
which is only about 30 MeV above the nucleon mass obtained in the
dynamical 3-body calculations of Ref.\cite{GL96b}. There the
shifted Gaussian of Eq. (25) results from a pure phenomenological fit.
\section{Results}
We diagonalize the Hamiltonian (22) in the six-quark basis (7-15) and calculate 
the $NN$ interaction potential in the Born-Oppenheimer approximation 
\begin{equation}
$$V_{NN}\left(Z\right) = \langle H\rangle_Z - \langle H\rangle_{\infty}\ ,$$
\end{equation}
where  $\langle H\rangle_Z$ is the lowest expectation value
obtained from the diagonalization at a given $Z$ and $\langle H\rangle_{\infty}
= 2m_N$ is the energy (mass) of two well separated nucleons. Here we
study the case $Z = 0$, relevant for short separation distances between the
nucleons. In Tables I and II we present our results for $IS$ = (01) and (10)
respectively, obtained from the diagonalization of $H$. From the diagonal
matrix elements $H_{ii}$ as well as from the eigenvalues, the quantity $2m_N$ =
1939 MeV has been subtracted according to (27). Here $m_N$ is the nucleon mass
calculated also variationally, with an $s^3$ configuration,
as mentioned at the end of the previous section. This value is
obtained for a harmonic oscillator parameter $\beta$ = 0.437 fm
\cite{PE97}. For sake of comparison with Ref. \cite{ST97} we take same value 
of $\beta$ for the six-quark system as well.\par
In both $IS$=(01) and (10) cases the effect of using
molecular orbitals is rather remarkable in lowering the ground state
energy as compared to the cluster model value obtained in the
four dimensional basis (16)-(19). Accordingly, the height of
the repulsive core in the $^1S_3$ channel 
is reduced from 915 MeV in the cluster model basis (see Appendix)
to 718 MeV in the molecular orbital basis. In the $^1S_0$ channel
the reduction is from 1453 MeV to 1083 MeV.
Thus the molecular
orbital basis is much better, inasmuch as the same two single 
particle states, $s$ and $p$, are used in both bases. 
 
The previous study \cite{ST97},
performed in a cluster model basis indicated that the dominant configuration is
associated to the symmetry $[42]_O[51]_{FS}$. It is the case here too
and one can see from Tables I and II that
the diagonal matrix element $H_{ii}$ of the state
$|42^+[42]_O[51]_{FS} >$ is far the lowest one, so that this
state is
much more favoured
than $|33[42]_O[51]_{FS} >$ . As explained above, such a
state represents a configuration with
two quarks on the left and four on the right 
around the symmetry centre.
At $Z \rightarrow \infty$ its energy becomes infinite 
i.e. this state behaves as a hidden colour state (see e.g. Ref.
\cite{HA81}) and it decouples
from the ground state. But at $Z = 0$ it is the dominant component
of the lowest state  with a probability of
87 \% for $IS$ = (01) and 93 \% for $IS$ = (10). The next
important state is
$|33[42]_O[51]_{FS} >$ with a probability of 10 \% 
for $IS$ = (01) and 4 \% for $IS$ = (10).
The presence of this state will become more and more important
with increasing $Z$. Asymptotically this state corresponds to 
a cluster model state with three quarks on the left and three 
on the right of the symmetry centre .\par
To have a better understanding of the lowering of the six-quark energy we
present in Tables III and IV the separate contribution of the kinetic energy
$KE$, of the confinement $V_{conf}$ and of the GBE interaction $V_{\chi}$ to the
dominant state in the cluster model $|s^4p^2[42]_O[51]_{FS}\rangle$
result and the dominant state in the
molecular basis case respectively. Table III corresponds to the $^3S_1$
channel and Table IV to the $^1S_0$ channel. We can see that $V_{conf}$ does not
change much in passing from the cluster model to the molecular orbital
basis. The kinetic energy $KE$ is higher in the molecular orbital basis which 
is natural because the $s^2p^4$ and $p^6$ configurations contribute with higher
energies than $s^6$ and $s^4p^2$. Contrary, the contribution of the GBE
interaction
$V_{\chi}$ is lowered by several hundreds of MeV in both channels, so that
$E = KE + V_{conf} + V_{\chi}$
is substantially lowered in the molecular orbital basis. This shows
that the GBE interaction is more effective in the molecular orbital basis than
in the cluster model basis. Note that $E$
differs from the value of the
diagonal matrix elements of Tables I and II by the 
additional quantity $6m - 2m_N$, where $m = m_u = m_d$.\par
The practically identical confinement energy in both bases shows
that the amount of Van der Waals forces, as discussed in \cite{ST97},
remains the same. However, the soft attraction brought in by the
Van der Waals forces does not play an important role at short
distances and it should be removed in further studies at intermediate
distances. \par
For both $IS$ = (01) and (10) sectors we also 
searched for the minimum of $\langle
H\rangle_{Z=0}$ as a function of the oscillator parameter $\beta$. For $IS$ =
(01) the minimum of 572 MeV  
has been reached at $\beta$ = 0.547 fm. For $IS$ = (10) the minimum
of 715 MeV was obtained at 
$\beta$ = 0.608 fm. These values are larger than the value of $\beta$ = 0.437
fm associated to the nucleon, which is quite natural because a six-quark system
at equilibrium
is a more extended object.

\section{Summary and conclusions}
We have calculated the $NN$ interaction potential at zero separation
distance between nucleons by treating $NN$ as a six-quark system
in a constituent quark model where the quarks interact via Goldstone
boson (pseudoscalar meson) exchange. The orbital part of the 
six-quark states was constructed from molecular orbitals instead
of the commonly used cluster model single particle states. The
molecular orbitals posses the proper axially and reflectionally
symmetries and are thus physically more adequate than the cluster
model states. Due to their orthogonality property they are also
technically more convenient. Here we constructed molecular
orbitals from harmonic oscillator $s$ and $p$ states. 
Such molecular orbitals are a very good approximation \cite{RO87} to
the exact eigenstates of a "two-centre" oscillator, frequently used
in nuclear physics in the study of the nucleus-nucleus potential.
The problem of calculating an $NN$ potential is similar in many ways.\par
We have shown that the upper bound of the ground state energy, 
and hence the height of the repulsive core in the $NN$ potential,
is lowered by about 200 MeV in the $^3S_1$ channel and by about
400 MeV in the $^1S_0$ channel. Hence using molecular orbitals
is more efficient than working with a cluster model basis.
A repulsive core of several hundred MeV is still present in
both channels. Due to the specific flavour-spin symmetry of the
GBE interaction the molecular type component 
$| {42}^{+}[42]_O [51]_{FS} >$ becomes dominant at short
range which implies that the $NN$ relative motion S-wave function
has a node at short distance due to the presence of the 
configurations $s^4p^2$ and $s^2p^4$. The dominance of the $[51]_{FS}$
symmetry will reinforce the repulsion in dynamical calculations.
In fact, it
has been shown \cite{OK84} that the phase shift calculated
within the resonating group method with a 
pure $[51]_{FS}$ state
shows a behaviour typical for potentials
with a repulsive core. In OGE models this effect is absent 
because none of the $[42]_O$ states is dominant (see e.g. \cite{ST88}) .
Note also that the configurations $s^2p^4$ or $p^6$ introduced
through the molecular orbitals might have an influence on the momentum
distribution of the $NN$ system as was discussed, for example,
in \cite{KW95} within the chromodielectric model.\par
The following step will be to calculate the $NN$ potential at $Z \neq 0$.
The Yukawa potential tail in Eq. (25) will bring 
the required long-range attraction. It would be interesting to
find out the amount of middle-range attraction brought in by
two correlated or uncorrelated pion exchanges.


\section{Appendix}
Ref. \cite{ST97} presented results obtained from the diagonalization
in a 5-dimensional basis. For comparison, here we need to remove the 
5th basis vector which does not have a corresponding one in the
molecular basis. The results of the diagonalization in a 4-dimensional
basis are given in Tables V and VI for $IS=(01)$ and $(10)$ respectively.

\vspace{0.8cm}
\noindent
{\bf{Acknowledgements.}} We are very grateful to L. Wilets and L. Glozman for
several useful comments.
 
\vspace{0.8cm}
\begin{table}
{\caption[states]{\label{states} Results of the diagonalization of the
Hamiltonian (22)-(26) for
$IS$ = (01). Column 1 - basis states, column 2 - diagonal matrix elements (GeV),
column 3 - eigenvalues (GeV) in increasing order, column 4 - lowest state
amplitudes of components given in column 1. The results correspond to $\beta$ =
0.437 fm . The diagonal matrix elements $H_{ii}$ and
the eigenvalues are relative to 2 $m_{N}$ = 1939 MeV (see text) }}
\begin{tabular}{|c|c|c|c|}
\hline
State &$H_{ii}$ - 2 $m_N\,$ & Eigenvalues - 2 $m_N$ & Lowest state amplitudes\\
\hline
$|33[6]_O[33]_{FS} >$ & 2.616 & 0.718 & -0.04571\\
\hline
$|33[42]_O[33]_{FS} >$ & 3.778 & 1.667 & 0.02479\\
\hline
$|33[42]_O[51]_{FS} >$ & 1.615 & 1.784 & -0.31762\\
\hline
$|33[42]_O[411]_{FS} >$ & 2.797 & 2.309 & 0.04274\\
\hline
$|42^+[6]_O[33]_{FS} >$ & 3.062 & 2.742 & -0.07988\\
\hline
$|42^+[42]_O[33]_{FS} >$ & 2.433 & 2.784 & 0.12930\\
\hline
$|42^+[42]_O[51]_{FS} >$ & 0.850 & 3.500 & -0.93336\\
\hline
$|42^+[42]_O[411]_{FS} >$ & 3.665 & 3.752 & 0.00145\\
\hline
$|51^+[6]_O[33]_{FS} >$ & 2.910 & 4.470 & -0.01789\\
\hline
\end{tabular}
\end{table}
\begin{table}
{\caption[statesbis]{\label{statesbis} Same as Table I but for $IS$ = (10)}} 
\begin{tabular}{|c|c|c|c|}
\hline
State &$H_{ii}$ - 2 $m_N$\,& Eigenvalues - 2 $m_N$ & Lowest state amplitudes\\
\hline
$|33[6]_O[33]_{FS} >$ & 3.300 & 1.083 & -0.02976\\
\hline
$|33[42]_O[33]_{FS} >$ & 4.367 & 2.252 & 0.01846\\
\hline
$|33[42]_O[51]_{FS} >$ & 2.278 & 2.279 & -0.20460\\
\hline
$|33[42]_O[411]_{FS} >$ & 3.191 & 2.945 & -0.04729\\
\hline
$|42^+[6]_O[33]_{FS} >$ & 3.655 & 3.198 & -0.07215\\
\hline
$|42^+[42]_O[33]_{FS} >$ & 2.796 & 3.317 & 0.13207\\
\hline
$|42^+[42]_O[51]_{FS} >$ & 1.167 & 4.058 & -0.96531\\
\hline
$|42^+[42]_O[411]_{FS} >$ & 4.405 & 4.459 & -0.00081\\
\hline
$|51^+[6]_O[33]_{FS} >$ & 3.501 & 5.070 & -0.01416\\
\hline
\end{tabular}
\end{table}

\begin{table}
{\caption[statester]{\label{statester} Parts of the energy expectation values
(GeV) of the dominant $6q$ state in the cluster model and the molecular
orbital basis for $IS$ = (01)}}
\begin{tabular}{|c|c|c|}
\hline
Energy &Cluster model &Molecular orbital\\
& $|s^4p^2[42]_O[51]_{FS}\rangle$ & $|42^+[42]_O[51]_{FS}\rangle$\\
\hline
$KE$ & 2.840 & 3.139\\
\hline
$V_{conf}$ & 0.385 & 0.364\\
\hline
$V_{\chi}$ & -2.384 & -2.754\\
\hline
$E$ & 0.841 & 0.749\\
\hline
\end{tabular}
\end{table}

\begin{table}
{\caption[statesquater]{\label{statesquater} Same as Table III but for $IS$ =
(10)}}
\begin{tabular}{|c|c|c|}
\hline
Energy &Cluster model &Molecular orbital\\
& $|s^4p^2[42]_O[51]_{FS}\rangle$ & $|42^+[42]_O[51]_{FS}\rangle$\\
\hline
$KE$ & 2.840 & 3.139\\
\hline
$V_{conf}$ & 0.385 & 0.364\\
\hline
$V_{\chi}$ & -1.840 & -2.437\\
\hline
$E$ & 1.385 & 1.066\\
\hline
\end{tabular}
\end{table}

\begin{table}
{\caption[states]{\label{states} Results of the diagonalization of the
Hamiltonian (22)-(26) for
$IS$ = (01). Column 1 - basis states, column 2 - diagonal matrix elements (GeV),
column 3 - eigenvalues (GeV) in increasing order for a
4 x 4 matrix ,
column 4 - components of the lowest state.
The results correspond to $\beta$ =
0.437 fm . The diagonal matrix elements  and
the eigenvalues are relative to 2 $m_{N}$= 1939 MeV  }}
\begin{tabular}{|c|c|c|c|}
\hline
State & Diag. elem - 2 $m_N$ & Eigenvalues - 2 $m_N$ & Lowest
state amplitudes\\
\hline
$|s^6[6]_O[33]_{FS} >$ & 2.346 & 0.915 & -0.10686\\
\hline
$|s^4p^2[42]_O[33]_{FS} >$ & 2.824 & 1.922 & 0.08922\\
\hline
$|s^4p^2[42]_O[51]_{FS} >$ & 0.942 & 2.956 & -0.98854\\
\hline
$|s^4p^2[42]_O[411]_{FS} >$ & 2.949 & 3.268 & 0.05843\\
\hline
\end{tabular}
\end{table}

\begin{table}
{\caption[states]{\label{states} Same as Table V but for $IS$ = (10)}} 
\begin{tabular}{|c|c|c|c|}
\hline
State & Diag. elem - 2 $m_N$ & Eigenvalues - 2 $m_N$ & Lowest
state amplitudes\\
\hline
$|s^6[6]_O[33]_{FS} >$ & 2.990 & 1.453 & -0.10331\\
\hline
$|s^4p^2[42]_O[33]_{FS} >$ & 3.326 & 2.436 & 0.09371\\
\hline
$|s^4p^2[42]_O[51]_{FS} >$ & 1.486 & 3.557 & -0.98723\\
\hline
$|s^4p^2[42]_O[411]_{FS} >$ & 3.543 & 3.899 & -0.07694\\
\hline
\end{tabular}
\end{table}


\begin{references}
\bibitem{OK84} M. Oka and K. Yazaki, Int.Rev.Nucl.Phys., vol. 1 (Quarks and
Nuclei, ed. W. Weise), World Scientific, Singapore, p. 490 (1984)
\bibitem{MY88} F. Myhrer and J. Wroldsen, Rev.Mod.Phys. {\bf{60}} (1988) 629
\bibitem{SH89} K. Shimizu, Rep.Prog.Phys. {\bf{52}} (1989) 1
\bibitem{KU91} A.M. Kusainov, V.G. Neudatchin and I.T. Obukhovsky, Phys.Rev.
{\bf{C44}} (1991) 1343
\bibitem{ZH94} Z. Zhang, A. Faessler, U. Straub and L.Ya. Glozman, Nucl.Phys.
{\bf{A578}} (1994) 573; A. Valcarce, A. Buchman, F. Fernandez and A. Faessler,
Phys.Rev. {\bf{C50}} (1994) 2246
\bibitem{FU96} Y. Fujiwara, C. Nakamoto and Y. Suzuki, Phys.Rev.Lett. {\bf{76}}
(1996) 2242
\bibitem{ST97} Fl. Stancu, S. Pepin and L.Ya. Glozman, Phys.Rev. {\bf{C56}}
(1997) 2779
\bibitem{GL96a} L.Ya. Glozman and D.O. Riska, Phys. Rep. {\bf{268}} (1996) 263
\bibitem{GL96b} L.Ya. Glozman, Z. Papp and W. Plessas, Phys. Lett. {\bf{B381}}
(1996) 311
\bibitem{GL97} L.Ya. Glozman, Z. Papp, W. Plessas, K. Varga and R.F.
Wagenbrunn,
Nucl.Phys. {\bf{A623}} (1997) 90c
\bibitem{ST87} Fl. Stancu and L. Wilets, Phys.Rev. {\bf{C36}} (1987) 726
\bibitem{ST88} Fl. Stancu and L. Wilets, Phys.Rev. {\bf{C38}} (1988) 1145
\bibitem{KO94} W. Koepf, L. Wilets, S. Pepin and Fl. Stancu,
Phys. Rev. {\bf{C50}} (1994) 614
\bibitem{HA81} M. Harvey, Nucl. Phys. {\bf{A352}} (1981) 301 ;
ibid. {\bf{A481}} (1988) 834
\bibitem{ST96} Fl. Stancu, Group Theory in Subnuclear Physics, Clarendon Press,
Oxford, 1996, chapters 4 and 10
\bibitem{MATH} S. Wolfram, The Mathematica book, Wolfram Media/ Cambridge
University Press, Cambridge, 1996
\bibitem{GPVW98} L.Ya. Glozman, W. Plessas, K. Varga and R. Wagenbrunn,
Phys. Rev. {\bf{D58}} (1998), to be published, e-print hep-ph/9706507
\bibitem{PS98} S. Pepin and Fl. Stancu, Phys. Rev. {\bf{D57}} (1998) 4475
\bibitem{PE97} S. Pepin, Fl. Stancu, M. Genovese and J.-M. Richard, Phys.Lett.
{\bf{B393}} (1997) 119
\bibitem{RO87} D. Robson, Phys. Rev. {\bf{D35}} (1987) 1029
\bibitem{KW95} W. Koepf and L. Wilets, Phys. Rev. {\bf{C51}} (1995) 3445
\end{references}
\end{document}